\documentclass{sn-jnl}

\usepackage[T1]{fontenc}
\usepackage{lmodern}
\usepackage{geometry}
\usepackage{verbatim}
\usepackage{graphicx}
\usepackage{float}
\usepackage{subcaption}
\usepackage{booktabs, multirow}
\usepackage{adjustbox}

\usepackage{amsmath, amssymb, amsfonts, amsthm}
\usepackage{mathrsfs, physics}
\usepackage{siunitx}

\usepackage{algorithm}
\usepackage{algpseudocode}
\usepackage{listings}

\usepackage[siunitx]{circuitikz}
\usetikzlibrary{positioning, shapes.geometric, decorations.pathmorphing, calc}

\usepackage{tikz}
\usetikzlibrary{arrows.meta, positioning, calc, shapes.geometric, backgrounds, fit, decorations.pathreplacing}

\usepackage[numbers]{natbib}
\usepackage{hyperref}
\usepackage{xcolor}
\usepackage[title]{appendix}
\usepackage{lineno}
\usepackage{bm}

\theoremstyle{thmstyleone}

\theoremstyle{thmstyletwo}

\theoremstyle{thmstylethree}

\begin{document}
\newcommand{\ctrl}{\bm{\theta}} 
\newcommand{\pph}{\bm{\varphi}}
\newcommand{\epspert}{\varepsilon}
\newcommand{\epsreg}{\delta}

\title{Shaping Frequency-Bin Biphoton Correlations via Programmable Optical Frequency Combs}

\author[1,2]{\fnm{Usman} \sur{Khan}}\email{ukhan1@stevens.edu}
\author[1,2]{\fnm{Daniel} \sur{Lopez}}\email{dlopezma@stevens.edu}
\author[1,2]{\fnm{Venkat Aneesh} \sur{Padmasola}}\email{vpadmaso@stevens.edu}
\author[1,2]{\fnm{Nirmal} \sur{Dangi}}\email{ndangi1@stevens.edu}
\author[1,2]{\fnm{Santosh} \sur{Kumar}}\email{skumar5@stevens.edu}
\author[1,2]{\fnm{Yuping} \sur{Huang}}\email{yhuang5@stevens.edu}

\affil[1]{\orgdiv{Department of Physics},
          \orgname{Stevens Institute of Technology},
          \orgaddress{\street{1 Castle Point Terrace},
                      \city{Hoboken}, \postcode{07030},
                      \state{New Jersey}, \country{USA}}}
\affil[2]{\orgdiv{Center for Quantum Science and Engineering},
          \orgname{Stevens Institute of Technology},
          \orgaddress{\street{1 Castle Point Terrace},
                      \city{Hoboken}, \postcode{07030},
                      \state{New Jersey}, \country{USA}}}

\abstract{
Programmable shaping of photon-pair spectral correlations is important for frequency-bin quantum information processing. However, achieving this control remains challenging in nonlinear optical systems without reliable forward models. Here, we demonstrate model-free, hardware-in-the-loop shaping of the measured joint spectral intensity of a cascaded photon-pair source driven by a programmable optical frequency comb. A blockwise simultaneous perturbation stochastic approximation controller optimizes up to 22 pump, signal, and idler parameters. Each iteration requires only two accepted hardware evaluations, independent of the parameter count. The controller converges to diverse target patterns on 3$\times$3, 4$\times$4, and 5$\times$5 frequency-bin grids. The optimized distributions achieve Pearson correlations between 0.66 and 0.98 and classical overlap fidelities between 0.32 and 0.62. Cross-target similarity analysis confirms that the optimized patterns remain target-specific. Source characterization yields a maximum coincidence-to-accidental ratio of approximately 585. Accuracy is limited by pump-phase transfer, parasitic nonlinear background, waveshaper constraints, amplitude-only filtering, and measurement noise. Joint spectral intensity contains no spectral-phase information. Therefore, the results demonstrate control of measured frequency-bin populations and correlations, not complete reconstruction of the two-photon state. Phase-sensitive characterization is required to verify inter-bin coherence and quantify the entanglement dimensionality of the output state. Our results establish direct measurement feedback as a practical approach to reconfigurable frequency-bin correlation engineering when an accurate end-to-end model is unavailable. 
}

\keywords{Hardware-in-the-loop optimization, Joint spectral intensity, Frequency-bin correlations, biphoton correlation shaping}

\maketitle

\section{Introduction}
\label{sec:intro}

Photonic systems are the leading platforms for quantum communication and information processing \cite{walmsley2015quantum, zhang2024entanglement}. Quantum information can be carried by a photon's polarization, arrival time, spatial path, or frequency. Frequency-bin encoding is attractive for three reasons. It travels well through standard telecom fiber. It is insensitive to fiber birefringence and interferometer drift. And it packs many quantum channels into a single spatial mode \cite{olislager2010frequency,lukens2016frequency,lu2023frequency,myilswamy2025chip}. These advantages have enabled high-dimensional entangled states in bulk and on-chip sources \cite{kues2017chip,reimer2016generation,kues2019quantum,imany201850,jaramillo2017persistent}, as well as practical frequency-bin quantum operations \cite{lu2018electro,lu2018controllable}.

Such states are most often produced by spontaneous parametric down-conversion (SPDC) in nonlinear media such as periodically poled lithium niobate (PPLN) waveguides \cite{caspani2017integrated}. The two-photon state is described by the joint spectral amplitude (JSA) and is set by the pump spectrum and the phase-matching function \cite{grice1997spectral,u2004efficient,mosley2008heralded}. A coincidence measurement records the joint spectral intensity (JSI), the squared magnitude of the JSA \cite{gianani2020measuring,kaneda2020spectral}. Shaping this map means choosing which frequency-bin pairs the source populates, which is the key resource for qudit encoding and multiplexed quantum channels
\cite{lu2023frequency,brecht2014demonstration}.

Most spectral engineering to date occurs in the design stage. Established tools include pump-pulse choice, phase-matching and dispersion control, engineered crystals, cavities, or filters
\cite{grice1997spectral,u2004efficient,mosley2008heralded,branczyk2010engineered,graffitti2018design,pomarico2012engineering}. These choices cannot be changed once the device is built. Programmable pulse shapers \cite{weiner2011ultrafast} add the missing flexibility. They have shaped entangled-photon spectra with programmable masks \cite{pe2005temporal,bernhard2013shaping,lukens2014orthogonal}, enabled arbitrary manipulation of the quantum waveform \cite{kowligy2014quantum} and electro-optic spectral-temporal control \cite{karpinski2017bandwidth,zhu2022spectral}. Comb-line amplitude and phase control have also been used for photonic neural-network computation on a related hardware platform \cite{lopez2026nonlinear}. Additional demonstrations include tailored nonlinear interactions \cite{ansari2018tailoring} and two-dimensional control of the joint spectrum \cite{shukhin2024two}. However, all of these operate in an open loop. The output is only as accurate as the model of the source. Closed-loop convergence to a user-chosen target JSI, directly on hardware, has not been demonstrated.

The difficulty is that in practice no accurate model exists. The measured JSI depends on comb generation, spectral shaping, the second-harmonic generation
(SHG), the SPDC, filtering, and detection. The combined system is too complex and drift-prone to model faithfully \cite{molesky2018inverse,carleo2019machine}.
Cascaded $\chi^{(2)}$ sources \cite{hunault2010generation,arahira2011generation} add four specific limitations. First, the SHG phase that pumps the SPDC stage is inherited from the shaped comb rather than set independently. Second, parasitic sum-frequency generation (SFG) between comblines adds background to the coincidence pattern. Third, a single waveshaper shared between the pump and the
photon-pair channels limits the speed and flexibility of the control loop. Finally, its finite spectral resolution sets the smallest feature that can be
programmed \cite{weiner2011ultrafast,roelens2008dispersion}. Together, these constraints limit which JSI patterns are physically reachable.

We therefore treat the entire apparatus as a black box. The control settings are learned directly from measured coincidences \cite{buckley2023photonic}. We use a blockwise simultaneous perturbation stochastic approximation (Block-SPSA) optimizer \cite{spall1992multivariate}. It requires only two hardware measurements per iteration, regardless of the number of parameters. Such efficiency is essential because each measurement is limited by the superconducting-nanowire detector (SNSPD) acquisition time. In this work, we shape the JSI of a cascaded SHG--SPDC source driven by a programmable optical frequency comb. We use up to 22 controls: six pump comb-line amplitudes, six phases, and $N_s$ ($N_i$) independent amplitude masks on the signal (idler) channel, giving $N=18$, $20$ and $22$ for the $3\times3$, $4\times4$ and $5\times5$ grids. We first locate the reachable spectral region through phase-matching characterization. The optimizer then converges to diverse target geometries at each grid size, and a cross-target similarity analysis confirms that the optimized JSI patterns remain target-specific. We also analyze explicitly how the hardware limitations above bound the achieved fidelity, including why increasing the number of pump comb lines cannot by itself close the remaining gap. 

The approach carries over to any programmable photonic platform with a coincidence-based objective \cite{piggott2015inverse,liu2021tackling}. Moreover, the data logged during optimization can train a differentiable digital twin of the experiment. This is a first step toward physics-aware training, in which gradients are backpropagated through a hardware-validated virtual copy of the system \cite{wright2022deep,hughes2018training,molesky2018inverse}. This capability ultimately points toward dynamically reconfigurable quantum networks, adaptive high-dimensional cryptography, and on-demand entanglement distribution for distributed quantum computing.

\section{Biphoton State and Joint Spectral Intensity}
\label{sec:background}

The quantum state of a photon pair generated by SPDC is naturally described in the frequency domain. For a weakly pumped, non-degenerate source, the two-photon state can be written as
\begin{equation}
|\Psi\rangle = \iint d\omega_s \, d\omega_i \, f(\omega_s,\omega_i)\,
\hat{a}_s^\dagger(\omega_s)\hat{a}_i^\dagger(\omega_i)|0\rangle,
\label{eq:biphoton_state}
\end{equation}
where $\hat{a}_{\mu}^\dagger(\omega)$ creates a photon of angular frequency $\omega$ in mode $\mu\in\{s,i\}$, and $f(\omega_s,\omega_i)$ is the JSA of the photon pair. The JSA is the product of the pump-envelope function and the phase-matching response of the nonlinear interaction. Therefore, it compactly captures how energy conservation, pump structure, and material dispersion together define the emitted two-photon state \cite{grice1997spectral,u2004efficient,mosley2008heralded}.

The quantity accessible in standard coincidence measurements is the JSI,
\begin{equation}
J(\omega_s,\omega_i)=|f(\omega_s,\omega_i)|^2,
\label{eq:jsi}
\end{equation}
obtained by spectrally resolving the signal and idler photons and recording the corresponding coincidence histogram \cite{gianani2020measuring, kaneda2020spectral}. The JSI shows how coincidence weight is distributed across the signal-idler frequency grid. It is the primary observable for characterizing photon-pair sources. For a source intended for frequency-bin encoding, shaping the JSI amounts to choosing which frequency-bin pairs the source populates. This property determines the usable qudit structure and channel allocation in downstream protocols \cite{olislager2010frequency, lukens2016frequency, lu2023frequency}.

The modal structure underlying these correlations is formalized by the Schmidt decomposition of the JSA \cite{law2000continuous},
\begin{equation}
f(\omega_s,\omega_i)=\sum_k \sqrt{\lambda_k}\,\psi_k(\omega_s)\chi_k(\omega_i),
\label{eq:schmidt}
\end{equation}
with nonnegative coefficients satisfying $\sum_k \lambda_k=1$. The Schmidt number ($K = \Big( \sum_k \lambda_k^2 \Big)^{-1}$) counts the effective number of jointly occupied spectral modes \cite{law2000continuous,brecht2014demonstration}. A value of $K \approx 1$ indicates a state that is nearly separable, while a larger value of $K$ signals an increase in multimode entanglement. Characteristic JSI shapes often carry qualitative signatures of this structure. A round, weakly correlated JSI suggests a nearly factorable state. A narrow anti-diagonal ridge reflects strong energy anticorrelation. A multi-lobed or comb-like pattern points to the occupation of multiple discrete frequency-bin pairs \cite{mosley2008heralded,kues2017chip}.

An important scope limitation frames the claims of this work. The JSI of Eq.~(\ref{eq:jsi}) fixes only the moduli $|f(\omega_s,\omega_i)|^2$, i.e., the population pattern in the frequency-bin basis. It carries no information about the relative phases of $f$. The Schmidt coefficients $\lambda_k$, and hence $K$ or any entanglement measure, therefore cannot be recovered from the JSI alone. Indeed, a separable and classically correlated mixture $\sum_{mn} |c_{mn}|^2\ket{m,n}\!\bra{m,n}$ reproduces any given JSI with zero entanglement. Accordingly, we restrict our quantitative claims to shaping the measured frequency-bin correlation structure. We do not certify an amount of entanglement from the JSI itself. The nonclassical character of the source is instead established by its CAR, which far exceeds the classical bound of 2 set by the Cauchy--Schwarz inequality for classical fields
\cite{clauser1974experimental,kuzmich2003generation}. We report the measured value in Sec.~\ref{sec:results}, confirming that the engineered correlations originate from genuine photon pairs rather than classically correlated light. Full reconstruction of the JSA, and with it certification of the engineered entanglement dimensionality, requires a phase-resolved technique. Examples include stimulated-emission tomography \cite{liscidini2013stimulated}, and interferometric and coherence-sensitive schemes \cite{gianani2020measuring}. We identify this as the natural next step, but we do not pursue it here.

For the source considered in this work, the experimentally controlled parameters act on two physically distinct ingredients of Eq.~(\ref{eq:biphoton_state}). Line-by-line shaping of the pump comb sets the spectral envelope and relative phases of the field that drives the nonlinear process, and therefore acts on every frequency-bin pair jointly rather than on individual bins. The post-SPDC amplitude masks on the signal and idler bands weight the transmitted coincidence amplitudes across the selected frequency bins. Together, these controls determine the measured coincidence landscape on the signal-idler grid. However, as discussed in Sec.~\ref{sec:intro}, the end-to-end map from control settings to measured JSI is too complex and drift-prone to predict with the fidelity required for model-based optimization. This is precisely why we treat the JSI as a directly measured optimization target and shape it in hardware (Sec.~\ref{sec:results}).

\section{Hardware-in-the-Loop Optimization}
\label{sec:method}

As established in Sec.~\ref{sec:intro}, the cascaded SHG-SPDC apparatus does not admit an accurate and differentiable first-principles model. The coupled hardware effects listed there are difficult to calibrate. We therefore treat the entire apparatus as a black box hardware map $\mathcal{H}$ and optimize it directly from measured coincidence data.

\begin{figure*}[t]
    \centering
    \resizebox{0.95\textwidth}{!}{%
        \input{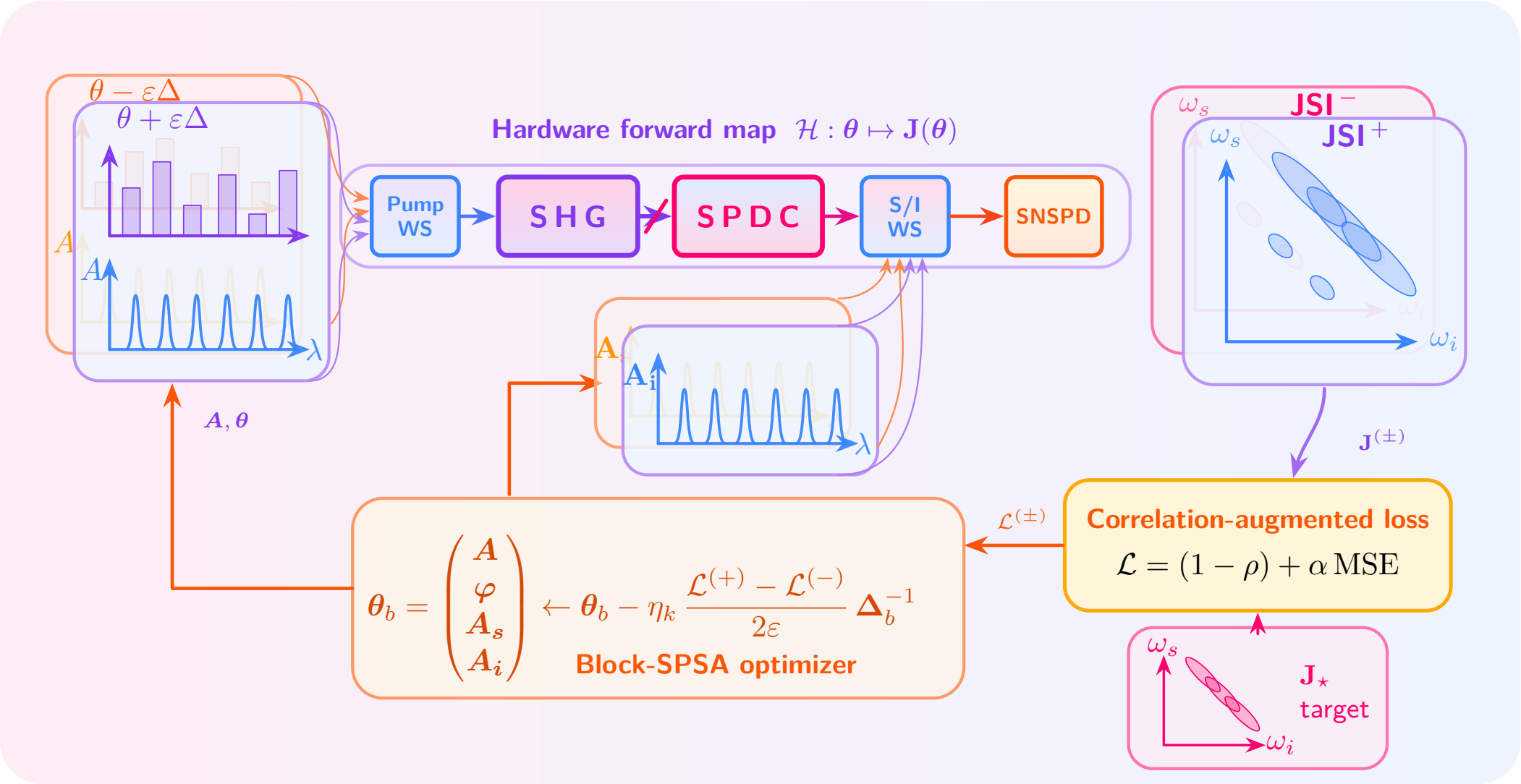}
    }
    \vspace{12pt}
    \caption{Closed-loop control framework for programmable biphoton frequency-bin correlations. The waveform parameters $\bm{\theta}$ are mapped by the nonlinear optical hardware to a measured JSI $\mathbf{J}(\bm{\theta})$, compared with a target $\mathbf{J}_\star$, and iteratively updated using a correlation-augmented block-SPSA optimizer.}
    \label{fig:sketch}
\end{figure*}

The closed-loop architecture is summarized in Fig.~\ref{fig:sketch}. The control vector $\bm{\theta}$ sets the line-by-line amplitude and phase mask $(\bm{A},\pph)$ applied to the pump comb, together with the post-SPDC amplitude masks $\mathbf{A}_s$ and $\mathbf{A}_i$ on the signal and idler bands. At each iteration the optimizer forms two perturbed configurations $\bm{\theta}\pm\varepsilon\bm{\Delta}$ that differ only in the currently active parameter block. Each configuration is sent to the hardware forward map $\mathcal{H}$: the shaped comb pumps SHG in Mixer~1, the second-harmonic field drives SPDC in Mixer~2, the generated pair is spectrally masked, and coincidences are recorded by the SNSPD. The two queries return
measured JSIs $\mathbf{J}^{(\pm)}$, which are compared with the user-specified target $\mathbf{J}_\star$ through the correlation-augmented loss $\mathcal{L}=(1-\rho)+\alpha\,\mathrm{MSE}$, yielding the scalar pair
$\mathcal{L}^{(\pm)}$. Their difference forms the block-restricted gradient estimate that updates only the active block before the cycle repeats. No step in the loop requires a forward model of the SHG--SPDC cascade; every evaluation is
supplied by the hardware itself.

\subsection{Problem Formulation}
\label{sec:problem}

Formally, let
\begin{equation}
\mathcal{H}:\bm{\theta}\mapsto \mathbf{J}(\bm{\theta})\in \mathbb{R}_{+}^{N_s\times N_i},
\label{eq:forwardmap}
\end{equation}
denote the experimental forward map. Here, $\pmb{\theta}\in\mathbb{R}^{N}$ is the control vector and $\mathbf{J}(\bm{\theta})$ is the JSI measured on a signal--idler frequency-bin grid $N_s\times N_i$. The analytic form of $\mathcal{H}$ is not assumed. The map is accessed only through experimental queries. A control vector is sent to the hardware, the apparatus is allowed to evolve physically, and a coincidence histogram is recorded.

The control vector contains four physically distinct parameter blocks
$\bm{\theta} = \left(\mathbf{u}_p, \mathbf{v}_p, \mathbf{u}_s, \mathbf{u}_i\right)$,
the unconstrained internal variables that map onto the physical masks
$(\bm{A}, \pph, \bm{A}_s, \bm{A}_i)$ of Fig.~\ref{fig:sketch} through the
reparametrization below. These are the pump comb-line amplitudes and phases, and signal--idler transmission weights. The total dimension is
\begin{equation}
  N  =  \underbrace{N_p}_{\text{pump ampl.}}  +  \underbrace{N_p}_{\text{pump phase}}  +  \underbrace{N_s}_{\text{signal ampl.}}  +  \underbrace{N_i}_{\text{idler ampl.}},
\end{equation}
where $N_p=6$ is the number of controlled pump comb-lines, spaced by 25 GHz and spanning $1549.13, 1549.33,\ldots,1550.13\,\mathrm{nm}$, that are selected from the programmable comb. The counts $N_s$, $N_i$ give the discretized signal and idler bins. The experiments reported here use an 18- to 22-dimensional control space.

The physical parameters are bounded. The optimizer therefore acts on unconstrained internal variables. These are mapped to admissible controls by $e = \sigma(u)$ with $\sigma(x) = (1+e^{-x})^{-1}$ for all amplitude-like controls, and by $\phi = \mathrm{wrap}(v) \in (-\pi,\pi]$ for the pump phases. This reparameterization keeps every transmission coefficient in $(0,1)$ and preserves the $2\pi$ periodicity of the phases in every hardware query. The search itself proceeds in an unconstrained space. Given a user-specified target $\mathbf{J}_\star$, the target is first normalized by its peak value $\mathbf{J}_\star \leftarrow \mathbf{J}_\star / \max(\mathbf{J}_\star)$. The objective is to match the correlation geometry, not an absolute coincidence rate.

Three structural features of this parametrization bound the reachable set. They partially explain the residual shape mismatch reported in Sec.~\ref{sec:results}. First, the post-SPDC signal and idler blocks act through amplitude masks only, $\mathbf{e} \in (0,1)^{N_s+N_i}$. They can attenuate frequency bins but cannot add power or impose relative phase. Post-SPDC shaping is therefore purely subtractive. Any target that requires more weight in a bin than the unshaped source supplies must be reached entirely through pump-side control. The achieved shaping also carries a coincidence-rate cost set by the masks. Second, only relative pump comb-line phases affect $\mathbf{J}(\bm{\theta})$. The global phase is unobservable, so the phase block carries $N_p - 1$ rather than $N_p$ effective
degrees of freedom. Third, pump-side control acts globally. A single comb line
does not address a single JSI bin. SHG mixes every pair of lines into the second
harmonic at $\omega_a + \omega_b$. SPDC then spreads each second-harmonic photon
across the whole contour $\omega_s + \omega_i = \omega_a + \omega_b$. Changing one pump amplitude or phase therefore redistributes weight along entire anti-diagonals of the JSI. Several line pairs also feed the same anti-diagonal coherently. The pump block thus sets the anti-diagonal populations collectively. The post-SPDC masks act locally, on individual rows and columns. This division of labor motivates the blockwise update of Sec.~\ref{sec:blockspsa}.

\subsection{Correlation-augmented loss function}
\label{sec:loss}

A practical difficulty in coincidence-based experiments is the slow drift of the total count rate, which is unavoidable here because a single optimization run acquires continuously for $1$--$19$\,hours. The drift arises from changes in coupling, temperature,
polarization, and detector efficiency. An objective based on mean-squared error (MSE) alone would read such global drifts as shape mismatch. We therefore combine a scale-insensitive correlation term with a pixelwise term. This makes the loop sensitive primarily to the geometry of the JSI.

Let $\mathbf{j}=\mathrm{vec}(\mathbf{J})$ and $\mathbf{j}_{\star}=\mathrm{vec}(\mathbf{J}_{\star})$ denote the vectorized measured and target JSIs. Let $\mathbf{a} = \mathbf{j} - \bar{\mathbf{j}}$ and
$\mathbf{b} = \mathbf{j}_\star - \bar{\mathbf{j}}_\star$ be their mean-removed versions. The Pearson correlation is 
\begin{equation}
\rho(\mathbf{a},\mathbf{b})=\frac{\mathbf{a}^{\mathsf{T}}\mathbf{b}}{\|\mathbf{a}\|_2\,\|\mathbf{b}\|_2+\epsilon_{o}},
\label{eq:pearson}
\end{equation}
with a small $\epsilon_{o}>0$ for numerical regularization. The optimization loss is
\begin{equation}
\mathcal{L}(\boldsymbol{\theta})
=
\bigl[1-\rho(\mathbf{a},\mathbf{b})\bigr]
+
\alpha\,\frac{1}{|\mathbf{a}|}\,
\|\mathbf{a}-\mathbf{b}\|_2^2,
\label{eq:loss}
\end{equation}
where $|\mathbf{a}|$ is the number of JSI pixels and $\alpha$ weights the MSE contribution. The correlation term rewards agreement in global geometry. It is largely insensitive to uniform rescaling of the counts. The MSE term suppresses trivial high-correlation solutions and enforces local fidelity. Without the correlation term, slow count-rate drift misleads the optimizer, and smeared or nearly uniform patterns can score deceptively well. The combination balances drift robustness against structural selectivity. The combined loss is designed to balance drift robustness and structural selectivity.

\subsection{Block-SPSA Optimizer}
\label{sec:blockspsa}

Since $\mathcal{H}$ is a physical apparatus, $\mathbf{J}(\bm{\theta})$ is not differentiable and the gradients cannot be backpropagated. Standard central finite differences over all $N$ controls require $2N$ hardware evaluations per full gradient. For the $5\times5$ grid, this corresponds to 44 evaluations. Each evaluation costs $75~\mathrm{s}$, so a single finite-difference gradient step would take $55~\mathrm{min}$ and a $260$-step optimization approximately $10$ days. We instead use Block-SPSA, which estimates a descent direction from only two evaluations per iteration regardless of $N$ \cite{spall1992multivariate,spall2005introduction}. The same $260$-step run
therefore completes in $10.8~\mathrm{h}$, a saving of a factor $N$, or
$18$--$22\times$ across the grid sizes studied here. The blockwise variant
corresponds to the physical structure of the control vector. More precisely, each iteration requires two accepted evaluations. Queries failing the coincidence gate below are repeated, so the realized cost and the run times quoted above are lower bounds. The blockwise structure follows the four physical parameter groups described below.

We partition $\bm{\theta}$ into four blocks
$\mathcal{B} = \{b_{\text{pump-amp}}, b_{\text{pump-phase}}, b_{\text{sig-amp}}, b_{\text{idl-amp}}\}$ --- the pump line amplitudes, pump line phases, signal-band amplitudes, and
idler-band amplitudes, respectively --- and cycle through them. In iteration $k$ with an active block $b$, a Rademacher perturbation
$\bm{\Delta}_b \in \{\pm 1\}^{|b|}$ is drawn and embedded in a sparse full-dimensional perturbation $\bm{\Delta}$. The hardware is
queried at $\bm{\theta}^{(\pm)} = \bm{\theta} \pm \varepsilon\bm{\Delta}$. The block-restricted gradient is estimated as
\begin{equation}
  \widehat{\mathbf{g}}_b
  =
  \frac{\mathcal{L}^{(+)} - \mathcal{L}^{(-)}}{2\varepsilon}
  \,\bm{\Delta}_b^{-1},
  \label{eq:spsa_grad}
\end{equation}
where $\bm{\Delta}_b^{-1}$ denotes the elementwise inverse. Only the active block is updated,
$\bm{\theta}_b \leftarrow \bm{\theta}_b -\eta_k\widehat{\mathbf{g}}_b$,
with a piecewise-decreasing learning rate starting at $\eta_0 = 0.2$. Restricting each perturbation to a single block (at most $N_p = 6$ dimensions) improves the quality of the gradient estimate relative to the full-dimensional SPSA. It also respects the distinct physical roles and the distinct noise sensitivities of the four parameter groups.

Two stabilization procedures constrain the search: spectral pre-alignment, which enforces reachability, and coincidence gating, which enforces gradient integrity.

\emph{Spectral pre-alignment}:
before the main loop, the signal-band center offset
$\delta\lambda_s\in\{-0.3,\ldots,0.3\}\,\mathrm{nm}$ is scanned. It is fixed at the value that minimizes the initial loss. This step addresses a reachability constraint, not a numerical convenience. The optimizer can shape the JSI only where the SPDC phase-matching function $|\phi(\omega_s,\omega_i)|^2$ has appreciable support. If the target frequency-bin grid is misaligned with the phase-matching ridge, the target JSI lies outside the reachable set of $\mathcal{H}$. No choice of $\boldsymbol{\theta}$ can then produce it. The pre-alignment registers the $N_s \times N_i$ grid inside the accessible region before optimization begins. The SFG map of Mixer~2 (Fig.~\ref{fig:pm}(b)) serves as a direct experimental diagnostic for this check. Some applications require JSI shaping at fixed wavelengths. Examples include interfacing with atomic memories, quantum nodes, or predetermined wavelength-division multiplexed (WDM) channels. In such cases, the offset cannot be freely chosen. The target window must instead be brought into the reachable set by other means: retuning the pump comb-lines (TLS center wavelength or OFCG line spacing), or by shifting the phase-matching ridge itself through temperature tuning, poling-period selection, or waveguide design.

\emph{Coincidence gating}: 
hardware queries whose total coincidence proxy
$C = \sum_{m,n}J_{mn}$ falls below a threshold $ C_{\min}$ are discarded and repeated. This prevents transient count dropouts and detector dead-time from corrupting the gradient estimates of Eq.~(\ref{eq:spsa_grad}). 

The procedure summarized in Fig.~\ref{fig:sketch} is stated in Algorithm~\ref{alg:bspsa}; we adopt a default budget of $T=260$ epochs with $\varepsilon=0.05$ and log the full control trajectory for post-analysis.

\begin{algorithm}[t]
\caption{Block-SPSA for hardware-in-the-loop JSI optimization}
\label{alg:bspsa}
\begin{algorithmic}[1]
\Require Target $\mathbf{J}_\star$; initial $\boldsymbol{\theta}^{(0)}$; epochs $T$; perturbation $\varepsilon$; rates $\{\eta_k\}$; threshold $C_{\min}$
\State Pre-alignment: scan $\delta\lambda_s$; fix $\delta\lambda_s^\star \leftarrow \arg\min \mathcal{L}$
\For{$k = 1,2,\dots,T$}
  \State $b \leftarrow \mathcal{B}\big[\,((k-1) \bmod 4)+1\,\big]$ \Comment{Cyclic block selection}
  \State Draw $\Delta_b \sim \mathrm{Rademacher}(|b|)$; embed into sparse $\Delta$
  \For{$s \in \{+1,-1\}$}
    \Repeat
      \State Query hardware at $\boldsymbol{\theta}^{(s)} = \boldsymbol{\theta} + s\varepsilon\Delta$; obtain $\mathbf{J}^{(s)}$; compute $C$
    \Until{$C \ge C_{\min}$}
    \State $\mathcal{L}^{(s)} \leftarrow \mathcal{L}(\boldsymbol{\theta}^{(s)})$
  \EndFor
  \State $\hat{g}_b \leftarrow \dfrac{\mathcal{L}^{(+)} - \mathcal{L}^{(-)}}{2\varepsilon}\,\Delta_b^{-1}$
  \State $\boldsymbol{\theta}_b \leftarrow \boldsymbol{\theta}_b - \eta_k \hat{g}_b$
  \State Apply sigmoid and phase-wrap to enforce physical constraints
  \State Log $b,\ \mathcal{L}^{(\pm)},\ \Delta_b,\ \hat{g}_b,\ \boldsymbol{\theta}$
\EndFor
\State \Return $\boldsymbol{\theta}^\star \leftarrow \boldsymbol{\theta}^{(T)}$
\end{algorithmic}
\end{algorithm}

\section{Experimental Setup}
\label{sec:experiment}

\begin{figure}[h]
  \centering
  \input{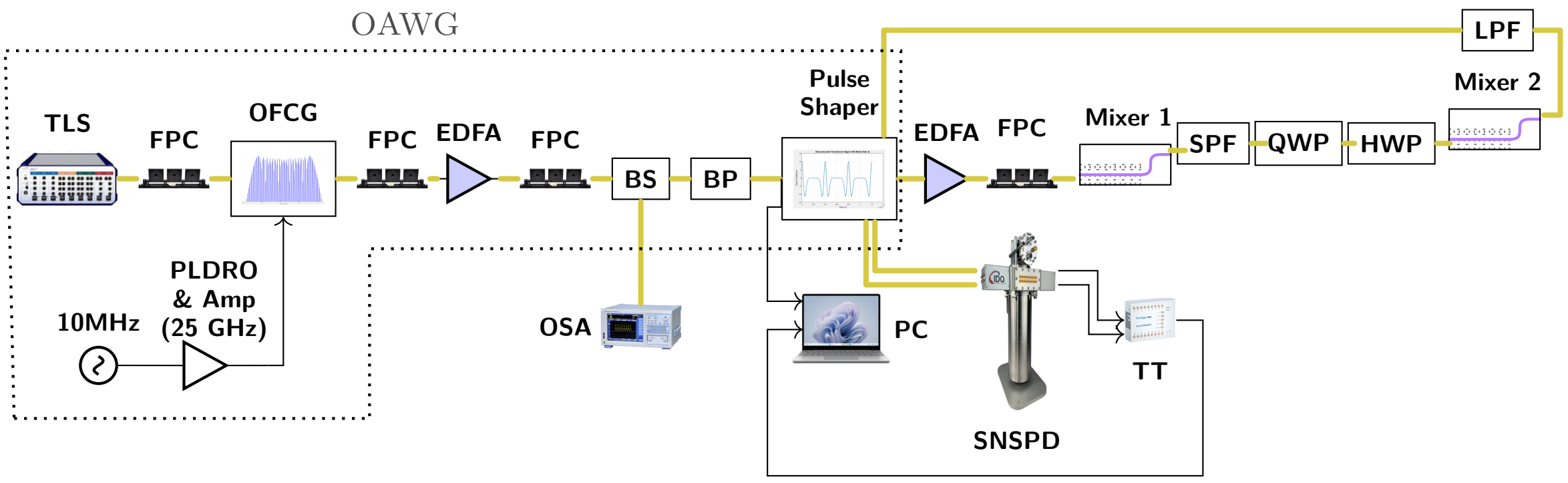}
  \caption{Hardware-in-the-loop experimental platform for programmable JSI shaping. 
  Solid yellow lines: optical paths; solid black lines: electronic control.}
  \label{fig:setup}
\end{figure}

The experimental platform realizes a programmable, measurement-driven SPDC source that supports frequency-bin entanglement. The closed loop shapes its measured JSI through optical waveform synthesis, nonlinear conversion, spectral filtering, and coincidence-resolved detection. As shown in Fig.~\ref{fig:setup}, the system is built around a programmable optical frequency comb, two cascaded PPLN waveguides (Mixer~1 and~2), and an SNSPD.

A tunable laser source (TLS) at $\sim\!1550\,\mathrm{nm}$ seeds an optical frequency comb generator (OFCG, WTEC-01-25) through a fiber polarization controller (FPC). The OFCG is temperature-stabilized at $50\,^\circ\mathrm{C}$. It uses phase modulation inside a Fabry--P\'erot cavity driven by a $25\,\mathrm{GHz}$ RF source. The result is a broadband comb with stable $25\,\mathrm{GHz}$ line spacing at $-12\,\mathrm{dBm}$ output power. The output combs are amplified by an erbium-doped fiber amplifier (EDFA, $23\,\mathrm{dB}$) and split by a 90:10 splitter. The $10\%$ arm continuously monitors the spectral stability on an optical spectrum analyzer (OSA). The $90\%$ arm passes through a $1.2\,\mathrm{nm}$ bandpass filter that selects six comb lines. The filtered comb lines then enter a programmable waveshaper (Finisar 16000A, $10\,\mathrm{GHz}$ resolution). The waveshaper applies line-by-line amplitude and phase control under MATLAB/Python automation (input channel~A, output channel~1). Together, the comb generator and waveshaper form the optical arbitrary waveform generation (OAWG) stage as shown in Fig.~\ref{fig:setup}.

The shaped pump is amplified by a second EDFA to compensate the waveshaper insertion loss. It is polarization-conditioned by in-line fiber polarizers and an FPC, then launched into the first $1\,\mathrm{cm}$ PPLN waveguide (Mixer~1), where SHG produces the ~$\sim$775\,nm field that pumps the next nonlinear stage. A short-pass filter (SPF) isolates this second-harmonic field from the residual telecom-band pump. A quarter-wave plate (QWP) and a half-wave plate (HWP) set the polarization of this field for phase-matched operation of the second stage. The filtered second-harmonic is injected into a second $1\,\mathrm{cm}$ PPLN waveguide (Mixer~2). There, nondegenerate SPDC generates signal--idler photon pairs in the telecom band. A long-pass filter (LPF) after Mixer~2 rejects the residual second-harmonic pump while transmitting the photon pairs. The nonlinear spectral responses of both waveguides are characterized independently (Sec.~\ref{sec:results}). These measurements anchor the placement of the frequency-bin grid within the phase-matched region. 

The generated photon pairs are routed back to the same waveshaper. It applies independent post-SPDC amplitude masks to the signal and idler channels on separate output ports. This shared waveshaper arrangement saves hardware. However, as noted in Sec.~\ref{sec:intro}, it couples the pump-shaping and pair-filtering functions of the control loop and constrains its reconfiguration speed and mask granularity. Moreover, the finite inter-port isolation of the waveshaper (approximately $-25~\mathrm{dB}$ crosstalk between channels) allows residual leakage between the pump, signal, and idler paths. To suppress this background, additional spectral filters are placed in both the signal and idler arms before the SNSPD. 

The masked photons are detected by the SNSPD. Coincidence events are time-tagged
and histogrammed by a time tagger (TT). This constructs the measured JSI,
$\mathbf{J}(\bm{\theta})$, for each hardware query in the optimization loop. The measured JSI is assembled on the $N_s \times N_i$ frequency-bin grid. Coincidences are accumulated over an integration time of $3~\mathrm{s}$ per
acquisition point within a coincidence window of $5~\mathrm{ns}$. These two
acquisition parameters set the photon-counting shot-noise level of the loss in
Eq.~(\ref{eq:loss}). They also give a quantitative meaning to the
coincidence-gate threshold $C_{\min}$ of Algorithm~\ref{alg:bspsa}. One hardware
query therefore requires $N_s N_i \times 3~\mathrm{s}$ of acquisition:
$27~\mathrm{s}$, $48~\mathrm{s}$ and $75~\mathrm{s}$ on the $3\times3$,
$4\times4$ and $5\times5$ grids respectively, plus waveshaper reconfiguration between points. A Block-SPSA iteration costs two accepted queries, so a $260$-epoch run spans approximately $3.9$, $6.9$ and $10.8~\mathrm{h}$. All reported runs therefore acquire data continuously over several hours, which sets the drift regime addressed by the loss. The source brightness and the coincidence to accidental ratio (CAR) measured under these settings are reported in Fig.~\ref{fig:CAR}.

\section{Results and Discussion}
\label{sec:results}

We evaluated the platform in four stages. First, we established the spectral operating window of the cascaded nonlinear source. Second, we set the pump regime from the measured tradeoff between pair flux and correlation quality. Third, we verified that the closed loop shapes the
measured JSI. Fourth, we quantified target matching and target specificity on $3\times3$, $4\times4$, and $5\times5$ frequency-bin grids.

\begin{figure}[h]
    \centering
    \resizebox{0.99\textwidth}{!}{%
        \input{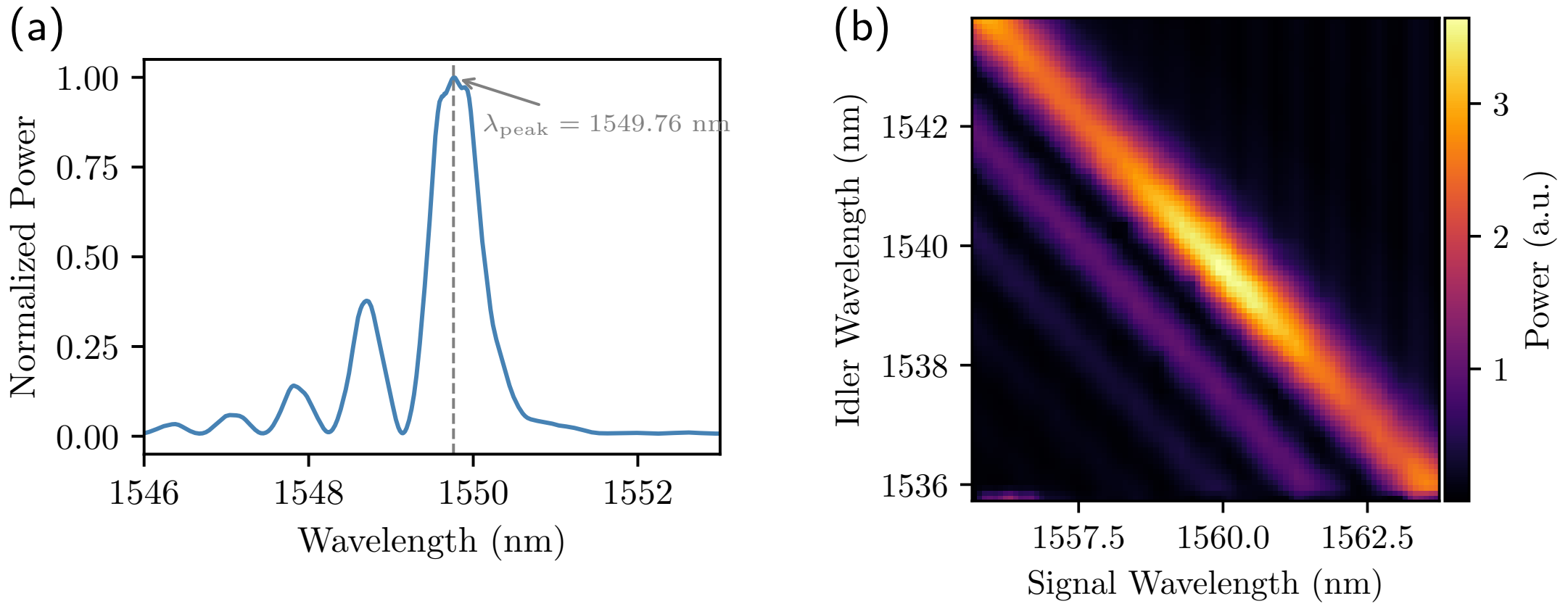}
    }
        \caption{Phase-matching characterization of the two PPLN waveguides. (a) SHG response of Mixer~1, defining the operating wavelength for visible-pump generation. (b) SFG map of Mixer~2, used as a proxy for the SPDC phase-matching function and to define the reachable signal--idler frequency region for JSI shaping.}
    \label{fig:pm}
\end{figure}

The two nonlinear stages were characterized separately. For Mixer~1, the SHG efficiency is measured against the telecom-band pump wavelength. As shown in Fig.~\ref{fig:pm}(a), the SHG phase-matching curve peaks at $\lambda_{\mathrm{peak}}=1549.76$~nm. This wavelength is chosen as the operating wavelength for efficient generation of the second-harmonic field that pumps SPDC in Mixer~2. For Mixer~2, we measured the SFG response against the signal and the idler wavelength. The SFG is the time-reversed counterpart of the SPDC. The measured two-dimensional SFG map shown in Fig.~\ref{fig:pm}(b) therefore provides a direct experimental proxy for the SPDC phase-matching function $|\phi(\omega_s,\omega_i)|^2$ \cite{kaneda2020spectral}. Its diagonal ridge traces the energy-conserving pairs $\omega_s+\omega_i=\omega_p$. The ridge width and extent define the phase-matching bandwidth and mark the region of the signal--idler plane where the discrete frequency-bin grid can be placed.

\begin{figure}[h]
  \centering
  \includegraphics[width=0.9\linewidth]{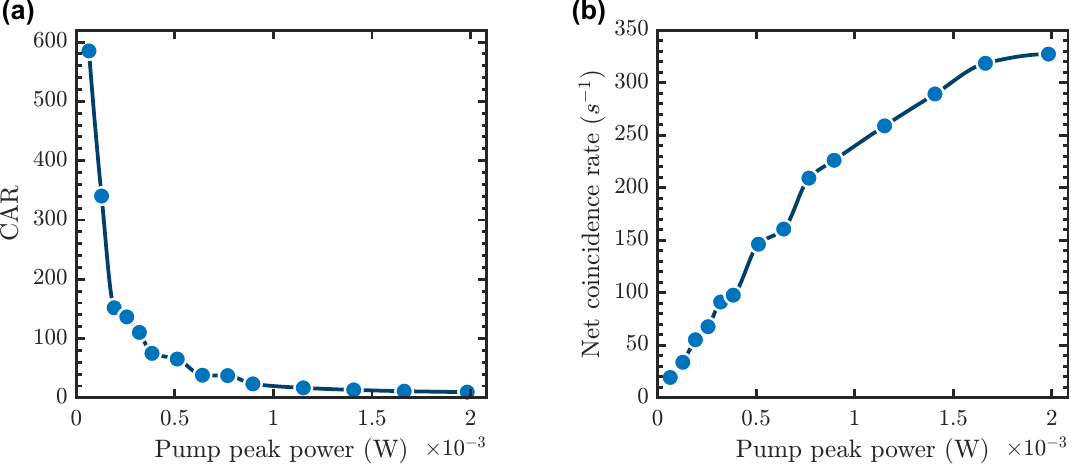}
    \caption{Source performance versus pump peak power. The data show the tradeoff between pair-correlation quality (a) and source brightness (b).}
  \label{fig:CAR}
\end{figure}

With the reachable region fixed, we characterized the source against the pump peak power as shown in Fig.~\ref{fig:CAR}. The CAR decreases monotonically with pump power, as shown in Fig.~\ref{fig:CAR}(a), reflecting the growth of accidental coincidences and multipair emission as pair generation is enhanced. By contrast, the net coincidence rate increases with the pump power, as shown in Fig.~\ref{fig:CAR}(b), reflecting the corresponding increase in source brightness. Together, these measurements define the tradeoff between correlation quality and pair flux. They therefore set the practical operating window for reliable JSI shaping. The CAR reaches a maximum of approximately $585$ at a pump peak power of $64~\mu\mathrm{W}$. This value indicates a low accidental-coincidence background and strong signal--idler pair correlations. CAR alone does not certify frequency-bin entanglement. We therefore use it to assess photon-pair source quality and select the operating power. The chosen operating point provides sufficient coincidence counts within the $3~\mathrm{s}$ integration time per point while limiting accidental and multipair contributions.

\begin{figure}[h]
  \centering
  \includegraphics[width=0.65\linewidth]{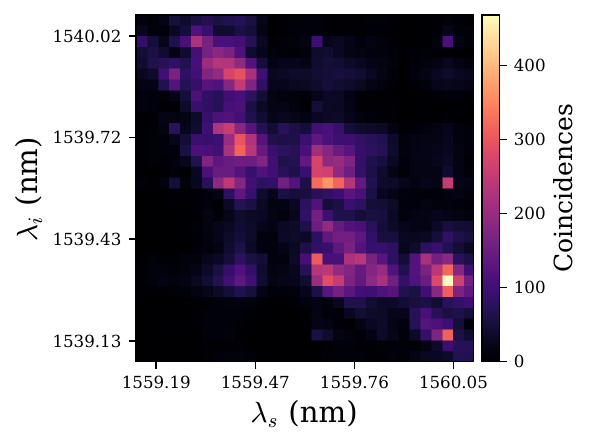}
  \caption{Measured JSI after Block-SPSA convergence. Coincidence weight is concentrated into discrete lobes along the energy-conserving contour, demonstrating programmable shaping of the biphoton spectral correlations from hardware feedback alone. Color scale: coincidence counts, \SI{3}{\second} integration per point.}

 \label{fig:jsi}
\end{figure}

We first establish that the closed loop shapes the measured JSI itself.
Figure~\ref{fig:jsi} shows the JSI after Block-SPSA convergence, assembled from
coincidences accumulated over \SI{3}{\second} per point within the
\SI{5}{\nano\second} window; no forward model of the SHG--SPDC cascade enters its
production. The coincidence weight is confined to compact, well-separated lobes, with peaks near 450 counts against a near-zero background, showing that pair emission has been concentrated into a discrete set of frequency-bin pairs rather than spread over the full phase-matching bandwidth. The lobes descend from upper left to lower right, so the occupied pairs lie along the energy-conserving contour
$\omega_s+\omega_i=\omega_p$ mapped independently by SFG in
Fig.~\ref{fig:pm}(b); the shaped structure therefore resides within the reachable region set by the phase-matching support. Together with the high CAR in Fig.~\ref{fig:CAR}, this establishes that line-by-line comb control and post-SPDC masking can impose a structured, spectrally anticorrelated population pattern in the measured biphoton JSI. Coincidence measurements provide the only feedback.

The map in Fig.~\ref{fig:jsi} also retains structure beyond the shaped pattern. Coincidence weight leaks into neighboring bins, and the bright features are unequal in amplitude. The origin of this residual maps onto the hardware limitations of Sec.~\ref{sec:intro}. The second-harmonic phase that pumps Mixer~2 is inherited from the shaped comb rather than imposed independently. This
residual pump-phase structure redistributes coincidence weight along the energy-conserving anti-diagonals, consistent with the broadening and the unequal peak amplitudes observed. Parasitic SFG among the pump comb lines contributes a weak, broadband pair background in nominally dark bins. The finite inter-port isolation of the shared waveshaper ($\approx-25$~dB crosstalk) adds a background
floor; the signal- and idler-arm filters (Sec.~\ref{sec:experiment}) suppress
but do not eliminate it. Finally, the $10$~GHz addressing resolution of the
waveshaper, together with its reduced accuracy for small amplitude increments,
limits precision at the finest programmed mask amplitudes. These are properties of the present hardware, not of the optimizer. The controller converges to the optimal pattern within this bounded reachable set.

\begin{figure}[h]
  \centering
  \includegraphics[width=\linewidth]{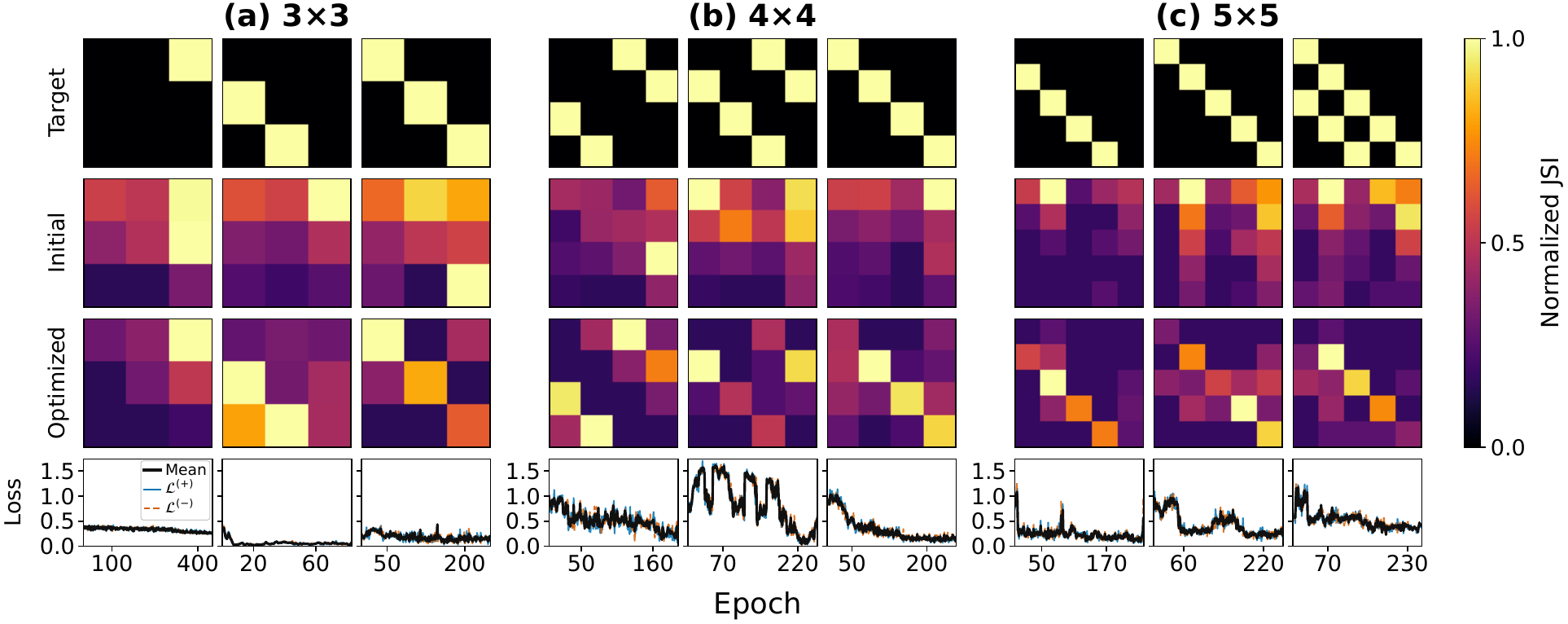}
    \caption{Representative Block-SPSA optimization results for (a)~$3\times3$, (b)~$4\times4$, and (c)~$5\times5$ frequency-bin grids, with three target geometries at each grid size. Rows, top to bottom: target JSI; initial (pre-optimization) measured JSI; measured JSI after Block-SPSA convergence; and loss trajectory. In the loss panels the solid black curve is the mean of the two SPSA evaluations $\mathcal{L}^{(\pm)}$, plotted as faint curves. Each JSI is normalized to its own maximum. Each column is an independent run, so the epoch axes differ between panels, with fluctuations characteristic of SPSA on noisy hardware.}
  \label{fig:representatives}
\end{figure}

\begin{figure}[h]
  \centering
  \includegraphics[width=\linewidth]{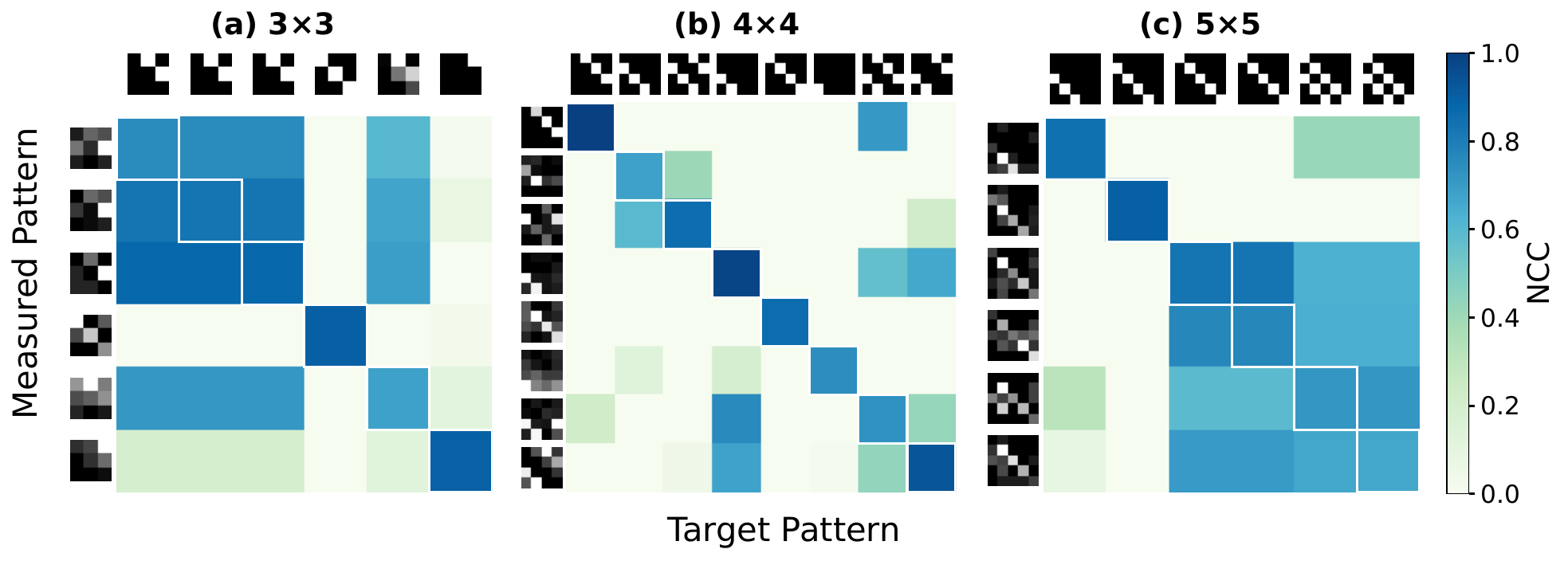}
    \caption{Cross-target similarity matrix for the optimized JSIs on the (a)~$3\times3$, (b)~$4\times4$, and (c)~$5\times5$ frequency-bin grids. Each entry gives the normalized cross-correlation between an optimized JSI and a target pattern. Diagonal entries compare each output with its intended target. Off-diagonal entries compare it with other targets. Lower off-diagonal values indicate stronger target specificity. Axis thumbnails show the optimized JSIs (rows) and target patterns (columns).}
  \label{fig:similarity}
\end{figure}

We next quantify how closely the optimized JSIs match the prescribed targets. Figure~\ref{fig:representatives} shows the target, initial, and optimized JSIs for three of the targets programmed at each grid size: (a)~$3\times3$, (b)~$4\times4$,
and (c)~$5\times5$. The full set enters the similarity analysis of
Fig.~\ref{fig:similarity}.
The targets span diagonal, anti-diagonal, block, and checkerboard geometries. The initial JSIs retain the unshaped source structure and deviate substantially from the targets. During optimization the feedback loop redistributes coincidence weight toward the programmed pattern, and the converged (measured) JSIs recover the dominant target features at every grid size. Residual macropixel-level deviations persist, caused by shot noise, finite coincidence counts, and slow experimental drift. Across these runs, the optimized JSIs achieve Pearson correlations $\rho=0.66$--$0.98$ and overlap fidelities $F=0.32$--$0.62$, with
\begin{equation}
  F \;=\;
  \frac{\left(\sum_{mn}\sqrt{J_{mn} J^{\star}_{mn}}\right)^{2}}
       {\sum_{mn} J_{mn}\,\sum_{mn} J^{\star}_{mn}},
  \label{eq:fidelity}
\end{equation}
the classical overlap between the normalized measured and target distributions. We report $\rho$ and $F$ together because they probe different failure modes. The correlation $\rho$ is offset- and scale-invariant, so it measures only whether high- and low-coincidence macropixels are placed correctly. The fidelity $F$ does
not have such an invariance. Although Eq.~\eqref{eq:fidelity} is normalized by the total weight of each distribution, it is not offset-invariant. Any coincidence weight in ideally-zero macropixels lowers $F$ even when the dominant pattern is recovered. In our data, the two are consistently
divergent, with high $\rho$ and lower $F$. We attribute this to a nonzero background of accidental coincidences in the measured maps, absent from the noiseless target. This floor is invisible to $\rho$ but is penalized by Eq.~\eqref{eq:fidelity}. We therefore read $\rho$ as the primary indicator of pattern reconstruction, and $F$ as a stricter benchmark that reflects the residual noise floor rather than a failure of pattern matching.

The loss panels (bottom row of Fig.~\ref{fig:representatives}) characterize the optimization itself. The solid black curve is the mean of the two SPSA evaluations $\mathcal{L}^{(+)}$ and $\mathcal{L}^{(-)}$, plotted as faint curves. The trajectories fluctuate but exhibit a clear downward trend, and both evaluations follow the same overall behavior. Each column is an independent run terminated at a loss plateau, so the epoch axes differ between panels. The sustained loss reduction confirms that SPSA extracts useful gradient information from only two hardware evaluations per iteration.

Finally, we examined whether the converged solutions are target-specific rather than generic. Figure~\ref{fig:similarity} shows the cross-target similarity matrix for each grid size: (a)~$3\times3$, (b)~$4\times4$, and (c)~$5\times5$. Each matrix element gives the normalized cross-correlation (NCC) between an optimized JSI, shown by a row, and a target pattern, shown by a column. Diagonal entries compare each optimized JSI with its intended target. Off-diagonal entries compare it with the other targets. The diagonal entries dominate in all three panels. The optimized outputs therefore remain target-specific and do not collapse onto a common pattern. Separation is cleanest for the $4\times4$ set, shown in Fig.~\ref{fig:similarity}(b), where the off-diagonal weight is uniformly low. The $3\times3$ and $5\times5$ sets in Fig.~\ref{fig:similarity}(a,c) retain some off-diagonal correlation, which appears between targets that share geometry, such as neighboring diagonal or checkerboard patterns. This is expected: geometrically similar targets yield similar measured JSIs, while dissimilar targets remain well separated.

Together these results define the demonstrated scope of the controller. Within the accessible source region it produces structured, target-specific JSIs
directly from hardware feedback, and it does so across grid sizes and target
geometries. The controller is model-free and requires no explicit description of
SHG efficiency, SPDC phase matching, or waveguide dispersion. It may therefore be
adapted to other programmable photonic platforms with coincidence-based
objectives. The demonstration nevertheless carries limitations that define its current operating range, discussed below. Addressing these constraints could extend the approach to adaptive quantum-source calibration, reconfigurable frequency-bin correlation engineering, and high-dimensional quantum photonic processing.

\subsection{Scope and Limitations of the HITL Approach}
\label{sec:hitl}

\emph{Shaping fidelity and controllability:} 
The optimizer reproduces the dominant structure of each target, but not every target is matched with high pointwise fidelity. Beyond the hardware error budget discussed above, two factors are intrinsic to the platform. First, post-SPDC control is amplitude-only, and therefore subtractive: bins can be attenuated but not amplified. The reachable set is thus bounded by the unshaped source spectrum and the available pump-side controls (Sec.~\ref{sec:method}). Second, the achievable JSI is confined to the support of the SPDC phase-matching function, which is why spectral pre-alignment is required (Sec.~\ref{sec:blockspsa}). Note that post-SPDC phase control would not relax either constraint: because the JSI is an intensity and no nonlinear process follows the SPDC stage, a phase applied to the signal or idler band leaves the measured coincidence pattern unchanged. Increased controllability must therefore come from the pump side or from the nonlinear response itself, which calls for a richer pump parametrization.

These constraints also determine how fidelity scales with the number of pump comb lines. Because pump control acts on whole anti-diagonals (Sec.~\ref{sec:problem}), once $N_p$ approaches the grid dimension every anti-diagonal is addressable and additional comb lines yield diminishing return. The post-SPDC masks act multiplicatively, producing a JSI that factorizes into an anti-diagonal term weighted by row and column transmissions, so targets requiring independent macropixel-level control remain unreachable regardless of comb-line count. Approaching unit fidelity therefore requires mode-mixing optics before detection or an engineered phase-matching response, not a larger comb. We regard the present amplitude-only result as a deliberate, hardware-limited first step.

\emph{Convergence speed:} SPSA gradient estimates are noisy. Each uses only two
evaluations corrupted by shot noise, so a large number of iterations is required.
For the 18- to 22-dimensional spaces studied here, individual runs reach a loss plateau within roughly $80$--$450$ epochs, depending on the target geometry and grid size. We adopt a default budget of $T=260$ epochs and extend it for targets that have not plateaued, which is why the epoch axes in
Fig.~\ref{fig:representatives} differ between panels. Convergence here means the loss reaching a plateau rather than a fixed stopping criterion. Larger control spaces or tighter fidelity targets will require a more sample-efficient optimizer. This motivates the physics-aware training direction below.

\emph{Characterization level:} We measure the JSI, which is an intensity and therefore phase-insensitive. We therefore characterize and shape frequency-bin correlations. The high CAR in Fig.~\ref{fig:CAR} indicates strong signal--idler pair correlations and a low accidental background. It is not an entanglement witness. We do not measure inter-bin coherence or certify entanglement dimensionality (Sec.~\ref{sec:background}).

\emph{Baselines:} We do not benchmark Block-SPSA against alternatives such as full-dimensional SPSA, Nelder--Mead, random search, or Bayesian optimization. The blockwise choice is motivated physically (Sec.~\ref{sec:blockspsa}) rather than by head-to-head comparison. A controlled baseline study at fixed target and measurement budget is a natural follow-up.

\section{Conclusion}
\label{sec:conclusion}

We demonstrated model-free, hardware-in-the-loop control of measured frequency-bin correlations in a cascaded SHG--SPDC photon-pair source. Block-SPSA optimized up to 22 pump, signal, and idler parameters directly from coincidence-resolved JSIs. Each block update required only two accepted hardware evaluations, independent of parameter count. A full central-difference gradient over 22 parameters would require 44 evaluations. For a 260-update budget, the nominal acquisition times are approximately ten days for central differences and $10.8$~h for Block-SPSA. This comparison concerns evaluation count and excludes waveshaper reconfiguration and repeated queries.

The measured phase-matching responses defined the accessible spectral region, while the selected pump power balanced photon-pair flux and correlation quality. The controller generated diverse patterns on $3\times3$, $4\times4$, and $5\times5$ frequency-bin grids. The optimized JSIs reached Pearson correlations $\rho=0.66$--$0.98$ and overlap fidelities $F=0.32$--$0.62$. Cross-target analysis confirmed target-specific outputs, and source characterization yielded a maximum CAR of approximately 585. Accuracy was limited by the phase-matching support, pump-phase transfer, parasitic nonlinear background, amplitude-only filtering, waveshaper constraints, and measurement noise. These effects reduced pointwise overlap but did not prevent recovery of the dominant target structures.

Additional pump comb lines cannot overcome the control limits of the present architecture. Higher fidelity requires mode-mixing or an engineered phase-matching response. Because the JSI contains no spectral-phase information, phase-sensitive characterization is required to quantify inter-bin coherence and entanglement dimensionality. Future work could combine these capabilities with independent pump and photon-pair shaping and more sample-efficient optimization. This could enable adaptive source calibration and reconfigurable high-dimensional quantum photonic processing. More broadly, this work establishes direct measurement feedback as a practical route to controlling complex photonic hardware without an accurate forward model.

\begin{backmatter}
\section*{Funding}
This research was supported in part by ACC-New Jersey under grant No.~W15QKN18D0040.

\section*{Disclosures}
The authors declare no competing interests.

\section*{Data availability}
Data are available from the corresponding author upon reasonable request.

\section*{Code availability}
The Block-SPSA optimization code will be made available at \url{https://github.com/daalopezm/JSI_machine_learning}.

\end{backmatter}

\bibliographystyle{bst/sn-mathphys-num}
\bibliography{references}

\end{document}